\documentclass[reprint,showpacs,showkeys,prl,aps]{revtex4-2}
\usepackage{graphicx}
\usepackage{subfig} 
\usepackage{epstopdf}
\usepackage{bm}
\usepackage{amsmath}

\usepackage{xcolor}
\begin{document}
\title{
Topological charge scaling at a quantum phase transition
}
\author{Piotr~Tomczak }
\affiliation{Faculty of Physics, ul.~Uniwersytetu~Pozna\'nskiego~2, 61-614 Pozna\'n, Poland}
\date{\today}

\begin{abstract}
We reexamine the Kosterlitz–Thouless phase transition in the ground state 
$|\Psi_0\rangle$ of an antiferromagnetic 
spin-$\frac{1}{2}$ Heisenberg 
chain with nearest and next-nearest-neighbor interactions $\lambda$
from a~different perspective: 
 After defining winding number (topological charge)
$W$ in the basis of resonating valence bond states, 
the finite-size scaling of
$\langle\Psi_0|W| \Psi_0 \rangle$,
$\langle\Psi_0|W|\partial_\lambda \Psi_0 \rangle$,
$\langle\partial_\lambda\Psi_0|W|\partial_\lambda \Psi_0 \rangle$
leads to 
the accurate value of critical coupling $\lambda_c=0.2412\pm0.0007$ 
and to the value of subleading critical exponent $\nu=2.000\pm0.001$.
This approach should be useful when examining the topological phase 
transitions in all systems described in the basis of resonating valence bonds.
\end{abstract}
\maketitle
{\em Introduction} --- At a quantum phase 
transition (QPT) properties of the ground state
of the quantum system change drastically due to quantum
fluctuations which are most clearly pronounced at zero temperature.
Although many approaches have been proposed to examine QPTs, to locate critical points, 
and to calculate the values of critical exponents,
an important question still remains:
{\em Is it possible to explore the critical behavior of a system at QPT by
examining the change of its ground state $|\Psi_0 \rangle$ in a critical region, 
especially when there is no possibility
to identify an order
parameter nor to establish a pattern related to symmetry breaking?}
Still, there exists a quest
for new approaches, based on scaling and renormalization to search and characterize QPTs.

Recently, mostly due to the interplay between information
theory and quantum many-body physics, new possibilities have emerged for the studying of QPTs. One of
the latest observations was that the fidelity, understood as
an overlap between the system ground states
calculated for the slightly shifted values of the parameter $\lambda$ 
whose change leads the system towards 
QPT, may be used to find it,
see e.g., Refs. [\onlinecite{Gu10}] and [\onlinecite{Wang15}].
Later on, this approach was extended to the fidelity susceptibility, 
$\chi_F = \langle \partial_\lambda \Psi_0 | \partial_\lambda \Psi_0 \rangle $,
and the phase transition was to be seen as a shift 
of~the~maximum in~the~dependencies of $f$ and $\chi_F$ on the parameter $\lambda$.
However, some difficulties were encountered with
finite-size scaling (FSS) of the fidelity susceptibility $\chi_F$
for topological QPTs:
It was unclear whether the maxima of $\chi_F$
obey FSS, 
and some attempts were made to interpret the emerging discrepancies\cite{Sun15} 
as logarithmic corrections to scaling.
It turned out recently\cite{Cincio}
that the maximum of $\chi_F$,
is shifted relative to the Kosterlitz-Thouless (KT) quantum critical point $\lambda_c$
by a~universal constant
$\frac{B^2}{36}$ towards the $ gapped $ phase in which
the correlation length $\xi$ falls exponentially,
$\xi(\lambda)\sim\mbox{exp}({B}/\sqrt{|\lambda-\lambda_c |})$. 
For this reason, 
the maximum in question does not scale with the system size $L$
as expected, see, e.g., Eq.~(5) in Ref.~[\onlinecite{Schwandt}]
or Eq.~(\ref{fidelity_scaling}) in what follows,
and the change of its position with the change of the system size $L$ cannot be used to find 
the critical properties of the system within FSS of fidelity
susceptibility.
On the other hand, the exponential decay of the correlation length
near the critical coupling $\lambda_c$ causes that the numerical
investigation of KT transition is hard,
since one has to examine rather large 
system (hundreds of spins) to avoid finite-size
effects.

In this paper, we address the problem of precise determination 
of the critical coupling and some critical exponents 
at a topological quantum phase transition and
show that the FSS method may be used for small systems
to locate such a quantum critical point and to determine
the critical exponents, despite the shift of the maximum mentioned 
earlier. 

{\em The model under consideration and its essential properties} ---
Our answer to the question posed in 
the {\em Introduction} is based on the 
reexamination of the quantum phase transition
in the known \cite{Eggert96,AffleckWhite92,Sandv10} one-dimensional 
spin-$\frac{1}{2}$ Heisenberg antiferromagnet with 
nearest neighbor interactions, set equal to 1, and next-nearest neighbor 
interactions, set  equal to $\lambda$.
The Hamiltonian reads
\begin{equation}
H = \sum_{\substack{i}}{\bf S}_i {\bf S}_{i+1}  +   \lambda \sum_{\substack{i}}{\bf S}_i {\bf S}_{i+2}.
\label{Hamiltonian}
\end{equation}
The ground state of this system depends on $\lambda$: for $\lambda<\lambda_c=0.241167$ 
it is similar to the ground state of the 1D antiferromagnet with nearest-neighbor interactions
(spin liquid phase), i.e.,
it is critical with the correlation function decaying 
in a power-like way, $\sim \sqrt{\mbox{log}(r)}/r$, $r$ stands for spin-spin separation.
Excitations are gapless - the finite-size triplet gap scales like $1/L$ 
(also with log correction).
For $\lambda>\lambda_c$, the system displays
a different ground state (dimerized phase): the correlation function decays exponentially,
$\sim \mbox{exp}[-r / \xi(\lambda)]$, with the distance,
and the triplet gap remains open in the thermodynamic limit.
The transition between these phases
is known to be of Kosterlitz-Thouless type.

{\em Resonating valence bond (RVB) basis and winding numbers} --- 
Using the RVB basis 
sheds additional light on the critical properties
of the considered system.
Recall then, briefly, the essential features
of RVB approach
\cite{Oguchi89, Sandv06}
to quantum  spin-$\frac{1}{2}$ systems:
Matrix elements of the Hamiltonian
are calculated not in the Ising basis
but in the (complete) nonorthogonal basis $|c_k\rangle$  taken from an overcomplete
set of
{\it singlet coverings}: 
\begin{equation}
\langle c_k|\,{\bf S}_i\cdot{\bf S}_j\,|c_l\rangle = 
(-1)^d \Big(\pm\frac{3}{4} \Big) \langle c_k|c_l \rangle,
\label{e4}
\end{equation}
$\langle c_k|c_l \rangle=2^{{\cal N}(c_k,c_l)-{\cal N}_s}$ 
with ${\cal N}(c_k,c_l)$ being the number of loops arising when the
coverings $\langle c_k|$ and $|c_l \rangle$  are drawn simultaneously
on the same lattice (``transition graph'') and ${\cal N}_s\!\!=\!\frac{L}{2} $ 
stands for the number of singlets in the system.
All singlets belonging to $|c_k\rangle$ are oriented;
$d$ denotes the number of {\em disoriented} ones 
one meets while moving along the loop in $\langle c_k|c_l \rangle$ containing $i$  and $j$.
Finally, $+\frac{3}{4}$ 
is taken if there is an even number of dimers between 
$i$ to $j$,  $-\frac{3}{4}$
in the opposite case.
To find the ground state of Hamiltonian (\ref{Hamiltonian}),
one solves the generalized 
eigenproblem: $H|\Psi_0 \rangle = E_0C |\Psi_0 \rangle$, 
with $C$ being the matrix formed from the scalar products $\langle c_k|c_l \rangle$.

\begin{figure}
\includegraphics[width=0.450\textwidth, angle=0]{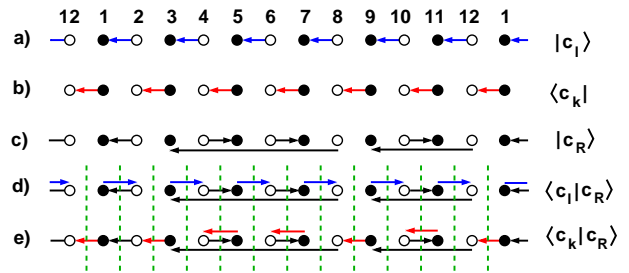}
\caption{Examples of calculating the winding numbers for two states $|c_l \rangle$ - blue (a)
and $|c_k \rangle$ - red (b) with respect to the reference state $|c_R \rangle$ - black (c) in a~12-spin system.
The system with periodic boundary conditions is 
represented as a line.
The winding number $W_{lR}=0$ is assigned to
the transition grph $\langle c_l |c_R\rangle$ (d)
since the number of singlets cut by green line
modulo~2 is equal to~0 (see text).
In Fig. e) the number of singlets modulo~2 crossed by green line
equals 1 and consequently $W_{kR}=1$.}
\label{coverings}
\end{figure}

The procedure of finding the ground state in the nonorthogonal
RVB basis has an advantage that can be seen 
when one uses periodic boundary conditions and maps the periodicity of the Hamiltonian onto a regular polygon.
The $L$ (even number) vertices of this polygon 
represent $L$ interacting Heisenberg spins-$\frac{1}{2}$ and
the $L$ edges represent interactions between spins.
Such a bipartite system of spins has $C_{L/2}^L/(\frac{L}{2}+1)$ linearly independent RVB basis states,
see, e.g., Ref. [\onlinecite{Oguchi89}],
and each of them may be characterized by its
{\em topological} winding number $W$ (topological charge).
To accomplish this,
one chooses any basis state  $|c_R\rangle$ as a~reference state, 
as noticed in Ref. [\onlinecite{Sandv11}] (see definition on page 3) and
obtains the winding number assigned to any other basis state  $|c_l\rangle $ 
from the transition graph $\langle c_l |c_R\rangle$ [\onlinecite{Sandv11}]
in the following way.
The polygon, representing the spin system
(vertices - spins, edges - interactions between them)
divides the plane into two disjoint areas;
one draws lines (green-dashed in Fig.~\ref{coverings})
connecting these areas and crossing each edge of the polygon.
Eventually, after drawing the transition graph $\langle c_l |c_R\rangle$ on the polygon, one counts how 
many singlets are cut by the green-dashed lines
connecting the inside and outside of the polygon
(see Fig.~\ref{coverings}).
Note that the number of singlets cut by $any$ green-dashed line
modulo~2 equals 0 for  Fig.~\ref{coverings}d and 1 for Fig.~\ref{coverings}e.
We, therefore, assign $W_{lR} = 0$ and $W_{kR} = 1$ to the basis states $|c_l\rangle $ 
and $|c_k\rangle $ with respect to the state $|c_R\rangle$.
Taking other basis states as reference states
and carrying out the similar procedure,
one finds the matrix of winding numbers $W_{pq}$ between all basis states.

The set of all basis states, 
after applying this definition, 
splits into two
disjoint subsets (sectors) $A$ and $B$ regardless of the choice of the reference state $|c_R\rangle$:
Each of the linearly independent basis states belongs to one of them.
For indices which number basis states 
within one sector it holds $W_{\alpha\alpha'}=W_{\beta \beta'}=0$,
while for different sectors one has $W_{\alpha\beta}=W_{\beta \alpha}=1$
for $\alpha,  \alpha'\in A$ and $\beta, \beta'\in B$.
This division leads to the block antidiagonal form of the matrix $W_{kl}$,
see an example for $L=8$ in Supplemental Material.
The basis states
belonging to the same sector can be
transformed into each other by a sequence of {\em local} singlet moves.
The transition graph resulting from a {\em local} move
does not wind around the entire system with periodic boundary conditions.
The basis functions from different sectors are not topologically equivalent:
one can not pass from a basis state belonging to one topological sector to 
a basis state belonging to another sector by making only local singlet moves, 
at least one move is required for which the transition graph winds around the entire system.
It means that the resonances between basis states from
different sectors
extend throughout the whole system (they are {\em global} ones), whereas 
for basis states from the same sector
only {\em local} resonances exist.
The global resonances are not topologically equivalent to the
local ones.

In what follows, we  apply the above observation to 
examine resonances present
in the ground state of the system under consideration
and show that their change from global to local
ones while changing $\lambda$ makes it possible to determine $\lambda_c$
and critical exponents after applying the FSS technique.
For this purpose, we consider how the following quantities 
depend on $\lambda$  for systems of various sizes $L$:\\
{\em i}) $\eta_T(\lambda) =\langle \Psi_0(\lambda)|W| \Psi_0(\lambda) \rangle $  (topological charge),\\
{\em ii}) $\chi_t(\lambda) = \frac{\chi_T(\lambda)}{L} =\frac{1}{L} \langle\partial_\lambda\Psi_0(\lambda)|W|\partial_\lambda \Psi_0(\lambda) \rangle$ (we will henceforth call it topological fidelity susceptibility), and\\
{\em iii}) $  \beta_t(\lambda) = \frac{\beta_T(\lambda)}{L} =\frac{1}{L} \langle \Psi_0(\lambda)|W|\partial_\lambda \Psi_0(\lambda) \rangle $ (we will henceforth call it topological connection).

Mean value of the Hermitian operator $W$ (topological charge) informs us
how large is the component of the vector $W|\Psi\rangle$ along $|\Psi\rangle$, 
i.e., what part of the state $|\Psi\rangle$ is composed of the basis functions belonging 
to different topological sectors.
If there were present only local (global) resonances in this state, this value would be 0 (1).
If, however,  local and global resonances are present, this value
will be between 0 and 1.
Similarly, 
the value of $\chi_T(\lambda)$
being the inner product between $|\Psi_\lambda\rangle$ and 
$W|\Psi_\lambda\rangle$,
tels us what amount of $W|\Psi_\lambda\rangle$ points in the direction of 
$|\Psi_\lambda\rangle$, i.e, what part of the fidelity susceptibility $\langle \partial_\lambda \Psi | \partial_\lambda \Psi \rangle $
is composed of the basis functions belonging to different topological sectors. 
Eventually, $\beta_t$ reveals what part of the change of $|\Psi\rangle$ 
with lambda
remaining parallel to $|\Psi\rangle$ itself, is composed of basis functions 
belonging to different topological sectors.
 Thus, the change of $\eta_T, \chi_T$, and $\beta_t$ with $\lambda$ enables the examination how
 the proportion changes between the basis states belonging to different topological sectors
 in a given state of the system.
This is equivalent to the change of the ratio between local and global resonances in this state.
As we will see further, $\eta_T, \chi_T$ and $\beta_t$ are subject to the 
finite-size scaling laws.

{\em Finite-size scaling of the topological charge and calculation of the critical coupling in the system} ---
Expressing the ground state in the RVB basis $|c_i\rangle$
and taking into account that for the system under
consideration all the coefficients $\alpha_i(\lambda)$ in this expansion may be chosen to be
{\em real numbers}, we find that
\begin{equation}
\eta_T(\lambda) = \sum_{k,l}\alpha_k (\lambda) \alpha_l (\lambda) W_{kl} \langle c_k |c_l \rangle.
\label{eta_T}
\end{equation}
Let us now assume that $\eta_T(\lambda, L)$ calculated for finite systems 
approaches similarly to
the spin stiffness (see Ref. [\onlinecite{Sandv10}], page~52 and Ref. [\onlinecite{Schu}]), 
its infinite size value with a~logarithmic size correction, i.e.,
\begin{equation}
\eta_T(\lambda_c, L) = \eta_T(\lambda_c, \infty)\Big(1+\frac{1}{2 \ln(L)+C}\Big),
\label{log_corr}
\end{equation}
and $C$ is a system dependent parameter.
\begin{figure}[!ht]
   \centering
   \subfloat{\includegraphics[width=0.25\textwidth, angle=0 ]{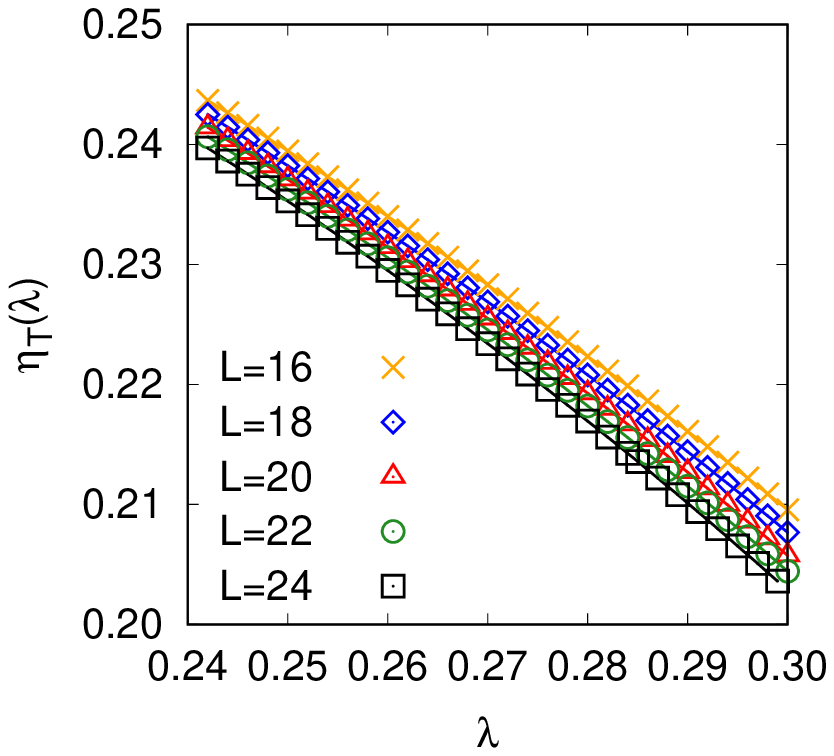}}
   \subfloat{\includegraphics[width=0.25\textwidth]{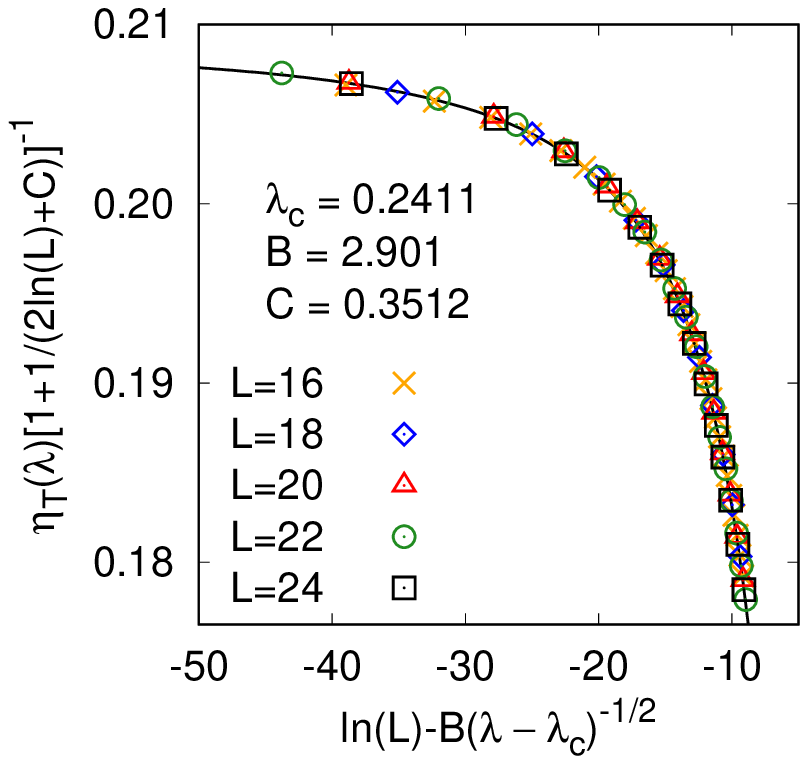}}
   \caption{Topological charge $\eta_T$ dependence on $\lambda$ above the transition
   $(\lambda_c=0.2411)$ before (left) 
   and after (right) rescaling, according to Eq.~(\ref{eta_scaling}), for systems up to 24 spins. }
   \label{ScalarProduct}
\end{figure}
Subsequently, 
using the FSS hypothesis and taking into account that above the transition
correlation length $\xi$ does not fall in a power-like manner, but rather exponentially
$[\xi(\lambda)\sim\mbox{exp}({B}/\sqrt{|\lambda-\lambda_c |})$],
leads [\onlinecite{Sandv10}] to the conclusion that
\begin{equation}
\eta_T(\lambda, L)\Big(1+\frac{1}{2 \ln(L)+C}\Big)^{-1} = 
\Lambda \Big(\ln(L)-\frac{B}{\sqrt{\lambda-\lambda_c}}\Big),
\label{eta_scaling}
\end{equation}
with $\Lambda$ being some universal scaling function.
In Fig.~\ref{ScalarProduct} it is shown that the scaling given by Eq.~(\ref{eta_scaling}) does occur:
we observe the collapse of $\eta_T(\lambda)$ calculated 
for different values of $L$ onto a single curve for $\lambda_c=0.2411$, $B=2.901$ and $C=0.3512$.
The value of $B$ is related to the KT transition width. The transition 
in the considered system is much $broader$ than that undergoing in the 
1D Bose-Hubbard model ($B = 0.261$), in XXZ spin-~$\frac{3}{2}$ model ($B = 1.61$) [\onlinecite{Cincio}]
or in 2D XY model ($B = 1.5$) [\onlinecite{Kosterlitz}]. 
This is also the reason why the maximum of $\chi_T$ is shifted
quite far from $\lambda_c$.

\begin{figure}[!ht]
   \centering
   \subfloat{\includegraphics[width=0.2\textwidth, angle=0 ]{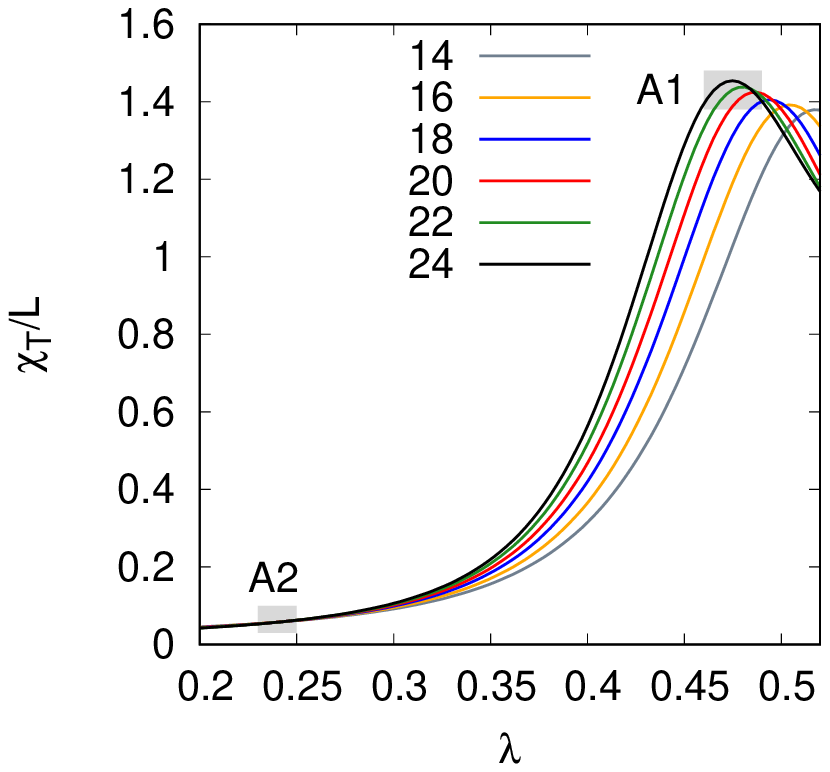}}
   \subfloat{\includegraphics[width=0.2\textwidth, angle=0]{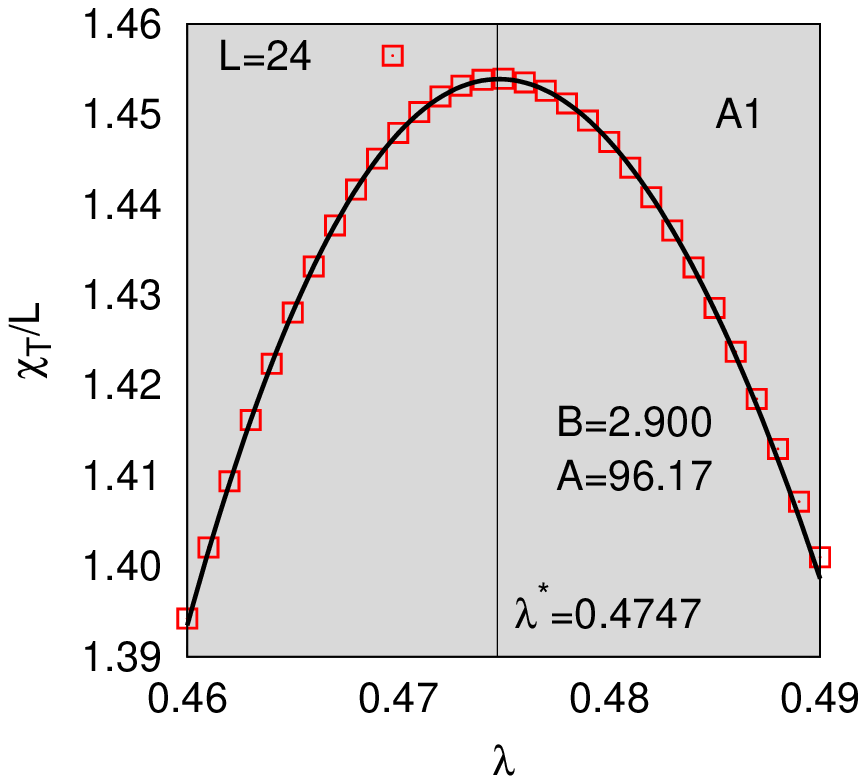}}\quad
    \subfloat{\includegraphics[width=0.20\textwidth]{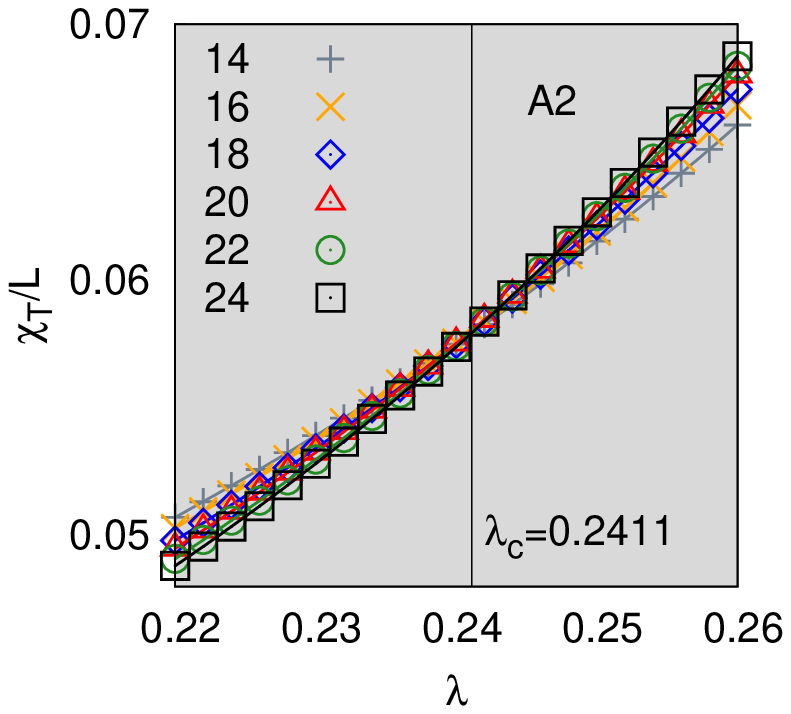}}
   \subfloat{\includegraphics[width=0.2\textwidth]{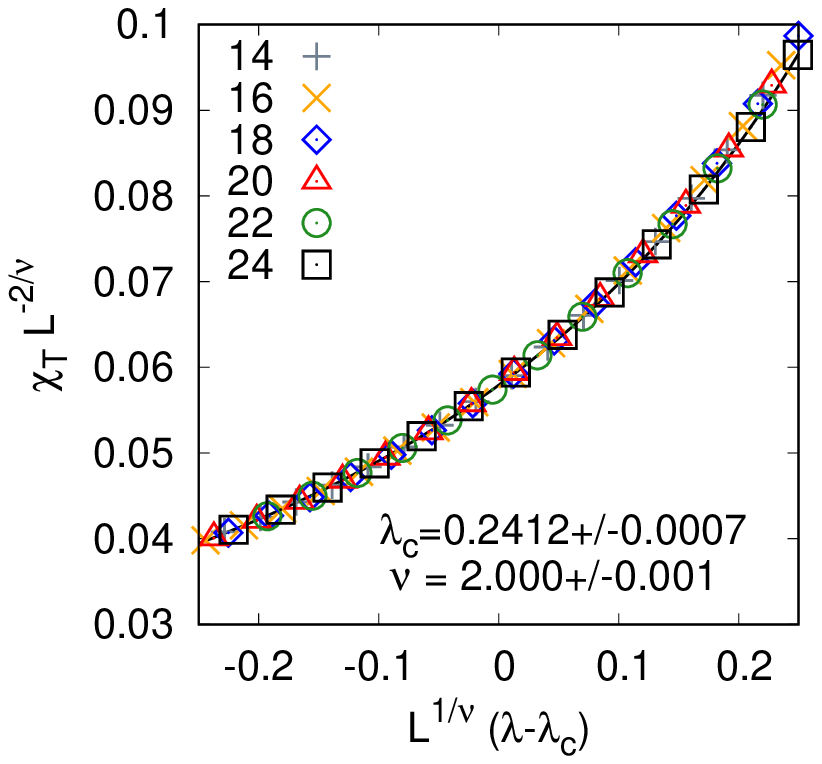}}
\caption{
Top left: Topological fidelity susceptibility for systems up to 24 spins versus $\lambda$.
Top right: enlarged maximum for 24 spins -- grey area A1.
Red squares - calculated values of $\chi_T/L$, black line - fit to the Eq.~(\ref{chi_shape}).
Bottom: Enlarged grey area A2 - topological
fidelity susceptibility before (left) and after rescaling (right)
for systems with $L=$16-24 spins.
The scaling collapse leads to the optimal values of $\lambda_c$ and exponent $\nu$.
The errors were estimated
by finding the collapse several times taking the numerical data with Gaussian noise 
with standard deviation equal to the accuracy 
of numerical differentiation while calculating $\chi_T$.
}
   \label{W0W1_dpdp}
\end{figure}

{\em Finite-size scaling of topological fidelity susceptibility} ---
One can calculate the topological fidelity susceptibility  
$\langle\partial_\lambda\Psi_0(\lambda)|W|\partial_\lambda \Psi_0(\lambda) \rangle$, from Eq.~(\ref{eta_T})
by replacing the coefficients $\alpha_k$ and $\alpha_l$ by their numerical derivatives with respect
to $\lambda$. Its dependence on $\lambda$ in systems up to $L=24$ spins
is shown in Fig.~\ref{W0W1_dpdp} (top left).
Notice the two gray areas in this figure marked by A1 and A2. 

The enlarged A1 area (right top of Fig.~\ref{W0W1_dpdp})
shows the maximum of $\chi_T/L$ for 24 spins.
From the position of this maximum, we can
independently determine the numerical value of parameter $B$
for the second time using the main result 
of Ref.~[\onlinecite{Cincio}]: The maximum of fidelity susceptibility $\chi_F$ should be
shifted with respect to $\lambda_c$ by the universal constant $\frac{B^2}{36}$.
To prove this, 
the authors of Ref.~[\onlinecite{Cincio}] assumed
that the singular part of the fidelity $ \langle \Psi_0(\lambda_1)| \Psi_0(\lambda_2) \rangle$
near the critical point
is a homogeneous function with respect to $\xi(\lambda_1)$ and $\xi(\lambda_2)$ and 
scales as\cite{Cincio}
$ \langle \Psi_0(\lambda_1)| \Psi_0(\lambda_2) \rangle = b^{-1} \Theta(\frac{\xi(\lambda_1)}{b}, \frac{\xi(\lambda_2)}{b} )$, with $\Theta$ 
being some universal function and $b$ -- a scaling factor.
From the definition of fidelity susceptibility, from this form of the scaling hypothesis, and from the exponential 
dependence of $\xi(\lambda)$, it follows that close to critical
point the fidelity susceptibility 
$\chi_F/L$ is given by
\begin{equation}
\chi_F(\lambda)/L = A \frac{\exp(({-B}/\sqrt{|\lambda-\lambda_c|}))}{|\lambda-\lambda_c|^3}.
\label{chi_shape}
\end{equation}
Let us now assume that $\chi_T$, being a part of $\chi_F$ is subjected to the same scaling.
Then we can find $B$ for the second time by fitting the calculated values of $\chi_T/L$ for 24 spins
near its maximum to
Eq.~(\ref{chi_shape}).
The continuous
line (in black) resulting from this fitting is shown on the right top of Fig.~\ref{W0W1_dpdp}.
The numerical value of $B=2.900$ determined by this method agrees well with the $B$ value obtained 
from topological charge scaling ($B=2.901$). The slight difference of the last digit may be due to the fact that the system of 24 spins was treated as if it were an infinite one. 

If we look more closely at the A2 area (left bottom of Fig.~\ref{W0W1_dpdp}),
we will see the crossing of $\chi_T/L$ calculated for systems from $L=14$ to $L=24$ spins
for some value of $\lambda_c$. To find this value and
the critical exponent $\nu$,  one takes into account
the argument that the topological fidelity susceptibility,
$\chi_T(\lambda)$, after appropriate rescaling 
of arguments and function values,
for different system sizes should collapse onto the same curve.
Let us assume  that the topological fidelity susceptibility $\chi_T(\lambda)$
at the critical point
scales as the fidelity susceptibility $\chi_F(\lambda)$\cite{Gu10,Polk15,Schwandt}
\begin{equation}
 \chi_T(\lambda)/L = L^{2/\nu-1}\Phi(L^{1/\nu}(\lambda - \lambda_c)),
\label{fidelity_scaling}
\end{equation}
with $\Phi$ being a universal scaling function.
The data from the left bottom of Fig.~\ref{W0W1_dpdp} has been plotted on the right side 
using rescaled values 
of each of the argument $L^{1/\nu}(\lambda - \lambda_c)$ and the function $\chi_F(\lambda)L^{-2/\nu}$.
Collapse occurs for $\lambda_c=0.2412\pm0.0002$ and $\nu=2.000\pm0.001$.
The singular part of $\chi_T(\lambda)$ should scale as $|\lambda-\lambda_c|^{\nu-2}$ [\onlinecite{Polk15,Schwandt}].
Since $\nu=2$, this scaling behaviour is subleading 
and the presence of $\chi_T$ peak is a result
of the dependence $\xi(\lambda)\sim\mbox{exp}({B}/\sqrt{|\lambda-\lambda_c |})$.
The value of $\nu = 2$ also results directly from Eq.(\ref{fidelity_scaling}), 
because then $\chi_T/L$ does not depend on $L$, which leads to the crossing of $\chi_T$ curves 
for different $L$ at the critical point. 

\begin{figure}[!ht]
   \centering 
   \subfloat{\includegraphics[width=0.22\textwidth, angle=0]{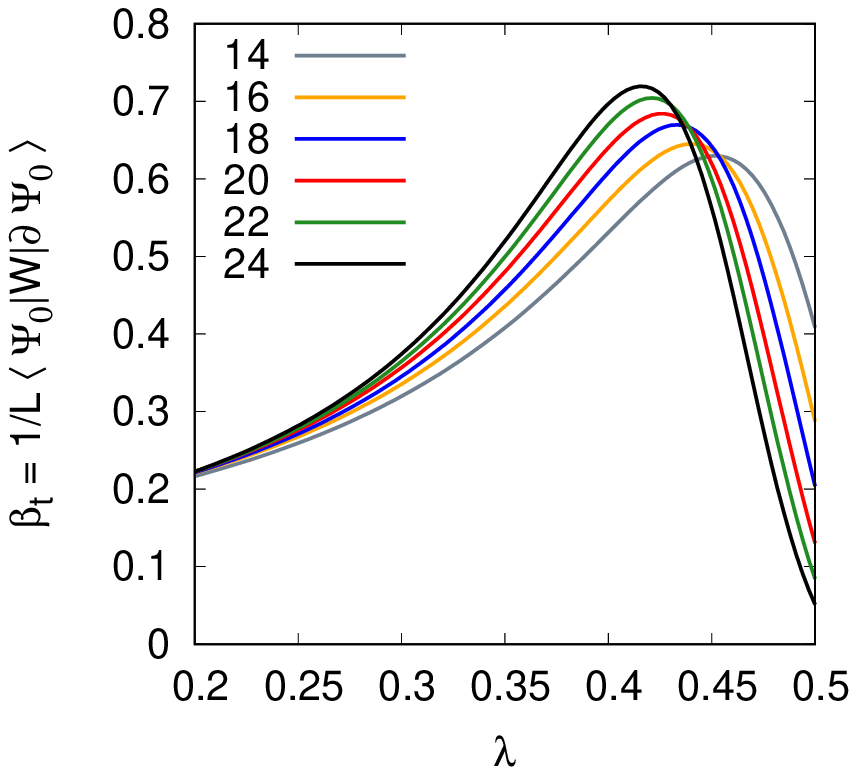}}\quad
   \subfloat{\includegraphics[width=0.22\textwidth, angle=0]{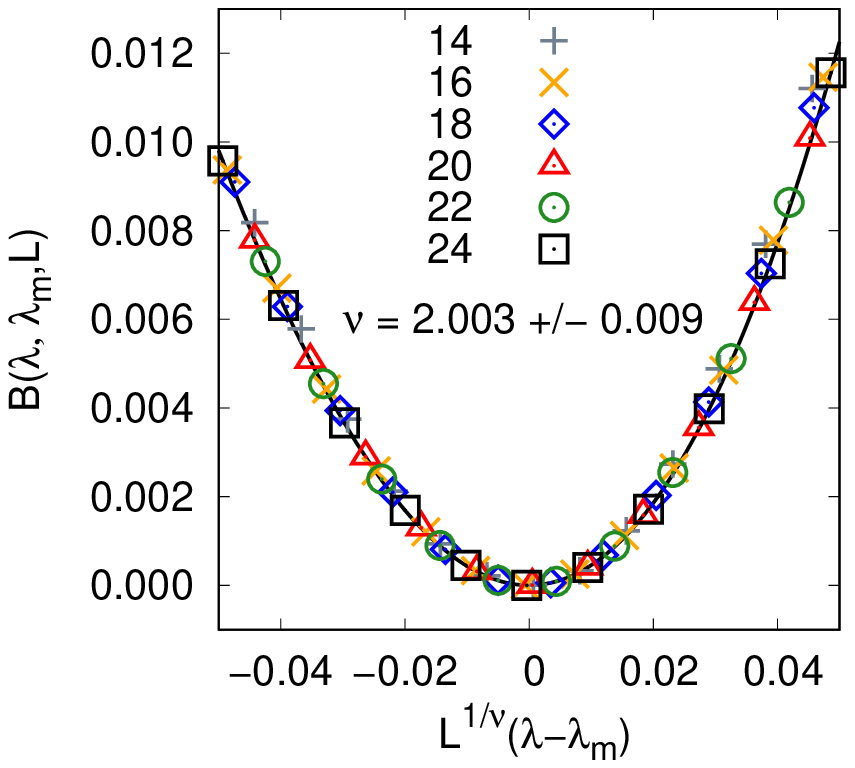}}
   \caption{
Left: The topological connection 
$\frac{1}{L}\langle \Psi_0(\lambda)|W|\partial_\lambda \Psi_0(\lambda) \rangle$
versus $\lambda$ before rescaling.
Right:
The value of the function $B(\lambda, \lambda_m, L)$ defined by Eq.~(\ref{ScalingFunction})
versus $L^{1/\nu}(\lambda-\lambda_m)$
for system sizes $L=$14-24.
As expected from the finite-size
scaling ansatz the data for different system sizes 
collapse on a single curve for $\nu=2.003$ (right). The error
is estimated as explained in the caption of Fig.~\ref{W0W1_dpdp}.
   }
  \label{scaling_connection}
\end{figure}

 {\em Topological connection scaling}
--- Let us now determine the subleading exponent 
$\nu$ for the second time from the FSS of the topological connection
$\frac{1}{L}\langle \Psi_0(\lambda)|W|\partial_\lambda \Psi_0(\lambda) \rangle=\frac{1}{L}\sum_{k,l}\alpha_k
\frac{\partial \alpha_l}{\partial \lambda} W_{kl} \langle c_k |c_l \rangle $.
The topological connection dependence on $\lambda$ displays a well-marked
peak for some $\lambda_m(L) > \lambda_c$. This maximum shifts  towards $\lambda_c$  with
increasing $L$, see Fig.~\ref{scaling_connection} (left),  and the value of $\beta_t(\lambda_m)$
 diverges logarithmically with $L$:
 $\beta_t(\lambda_m) \propto 0.170\,\mbox{log}L$.
 Similar dependence was reported\cite{Shi06} for the phase transition in the ground state 
of the quantum XY chain. This logarithmic divergence
suggests \cite{Barber} that 
it is possible to extract a critical exponent $\nu$
from the scaling of the function
\begin{equation}
\begin{split}
B(\lambda, \lambda_m, L) =  \Big(1 - \mbox{e}^{\beta_t(\lambda(L)) - \beta_t(\lambda_m(L))}\Big)
& \\ \propto L^{1/\nu}(\lambda(L)-\lambda_m(L)),
\label{ScalingFunction}
\end{split}
\end{equation}
$\lambda_m(L)$ stands for the peak position for a given
$L$ (Fig.~\ref{scaling_connection} - left) and $\beta_T(\lambda(L))$
- for the topological connection near $\lambda_c$ for a given $L$ . 
All data for systems with different $L$
collapse onto a single curve as shown in Fig.~\ref{scaling_connection} (right), 
after taking $\lambda_c=0.2411$ for $\nu=2.003\pm0.009$.

{\em Summary} ---
The topological charge $\eta_T$, the topological susceptibility
 $\chi_T$ and the topological connection $\beta_t$ obey the finite-size scaling.
 The scaling occurs for relatively small systems.
Treated as a test, made it possible to determine accurately the critical coupling $\lambda_c$
and subleading exponent $\nu$ in a well-known topological phase 
transition in spin-$\frac{1}{2}$ Heisenberg chain with nearest and next-nearest-neighbor interactions.
We hope that the presented results will stimulate further exploration
of topological critical phenomena in other {\em Resonating Valence Bond} 
systems in which there is no possibility
to define an order parameter.

\vspace{1cm}
\noindent
{\em Acknowledgments.}\!\! --- Author would like to thank Marcin Tomczak 
and to Piotr Jabło\'nski for stimulating discussions.
Numerical calculations were performed 
at Pozna\'n Supercomputing and Networking Center under Grant no. 284.


\pagebreak
\widetext
\begin{center}
\textbf{\large Supplemental Material: Topological charge scaling at a quantum phase transition}
\end{center}
\setcounter{equation}{0}
\setcounter{figure}{0}
\setcounter{table}{0}
\setcounter{page}{1}
\makeatletter
\renewcommand{\theequation}{S\arabic{equation}}
\renewcommand{\thefigure}{S\arabic{figure}}
\renewcommand{\bibnumfmt}[1]{[S#1]}
\renewcommand{\citenumfont}[1]{S#1}

Let us take into account, as an example, 
the 1D spin system consisting of $L=8$ spins with periodic boundary conditions. 
The complete, non-orthogonal RVB basis consists of 14 following 
basis functions (singlet coverings)
\begin{equation}
 \begin{split}
 |c_1 \rangle = (1, 8) (3, 2) (5, 4) (7, 6),\\      
 |c_2 \rangle = (1, 2) (3, 8) (5, 4) (7, 6),\\      
 |c_3 \rangle = (1, 6) (3, 2) (5, 4) (7, 8),\\      
 |c_4 \rangle = (1, 8) (3, 2) (5, 6) (7, 4),\\      
 |c_5 \rangle = (1, 8) (3, 4) (5, 2) (7, 6),\\      
 |c_6 \rangle = (1, 2) (3, 8) (5, 6) (7, 4),\\      
 |c_7 \rangle = (1, 6) (3, 4) (5, 2) (7, 8),\\      
 |c_8 \rangle = (1, 2) (3, 4) (5, 6) (7, 8),\\      
 |c_9 \rangle = (1, 2) (3, 4) (5, 8) (7, 6),\\      
 |c_{10} \rangle = (1, 2) (3, 6) (5, 4) (7, 8),\\      
 |c_{11} \rangle = (1, 4) (3, 2) (5, 6) (7, 8),\\      
 |c_{12} \rangle = (1, 8) (3, 4) (5, 6) (7, 2),\\      
 |c_{13} \rangle = (1, 4) (3, 2) (5, 8) (7, 6),\\      
 |c_{14} \rangle = (1, 8) (3, 6) (5, 4) (7, 2).\\ 
\end{split}
 \end{equation}

Coverings 1-7 and 8-14 belong to different topological sectors. To make it clear, let us consider singlet moves that need to be made to go, say, from $|c_1 \rangle$ to $|c_4 \rangle$ (same sector) 
and from $|c_1 \rangle$ to $|c_9 \rangle$ (between sectors). In the first case, it is enough to make 
one local  
singlet move
$(1, 8) (3, 2) {\color{red}(5, 4) (7, 6)} \to (1, 8) (3, 2) {\color{red}(5, 6) (7, 4)}$. 
By {\em local singlet move}, $|c \rangle \to |c^\prime \rangle$,
we understand such a rearrangement of two singlets in a given covering, that
any loop length $\mathcal S$ associated with the transition graph $\langle c |c^\prime \rangle$ is
less than $L$ (in the considered case $\mathcal S(4,5,6,7,4)=6\le L$.
In other words, any of resonances associated with {\em local move} do not 
cover the entire system.

In the second case ($|c_1 \rangle$ to $|c_9 \rangle$) two moves are needed 
${\color{red}(1, 8) (3, 2)} (5, 4) (7, 6) \to {\color{red}(1, 2) (3, 8)} (5, 4) (7, 6)$ 
and subsequently $(1, 2) {\color{red}(3, 8) (5, 4)} (7, 6) \to (1, 2){\color{red} (3, 4) (5, 8)} (7, 6)$.
The first move is a local one: $S(1,2,3,8,1)=6$, whereas the second move 
is a {\em global} one and  extends throughout the entire system: $S(3,4,5,8,3)=L=8$.

Therefore the transition from one the basis function to another within the same sector 
is associated  with local resonances, while
the transition from one topological sector to
the other requires a global resonance. 
We can also infer this from the form of the matrix $W_{kl}$,

\begin{equation}
W_{kl}(L=8)=\left[ \begin {array}{cccccccccccccc} 
                     0&0&0&0&0&0&0&1&1&1&1&1&1&1
\\ \noalign{\medskip}0&0&0&0&0&0&0&1&1&1&1&1&1&1
\\ \noalign{\medskip}0&0&0&0&0&0&0&1&1&1&1&1&1&1
\\ \noalign{\medskip}0&0&0&0&0&0&0&1&1&1&1&1&1&1
\\ \noalign{\medskip}0&0&0&0&0&0&0&1&1&1&1&1&1&1
\\ \noalign{\medskip}0&0&0&0&0&0&0&1&1&1&1&1&1&1
\\ \noalign{\medskip}0&0&0&0&0&0&0&1&1&1&1&1&1&1
\\ \noalign{\medskip}1&1&1&1&1&1&1&0&0&0&0&0&0&0\\ \noalign{\medskip}1
&1&1&1&1&1&1&0&0&0&0&0&0&0\\ \noalign{\medskip}1&1&1&1&1&1&1&0&0&0&0&0
&0&0\\ \noalign{\medskip}1&1&1&1&1&1&1&0&0&0&0&0&0&0
\\ \noalign{\medskip}1&1&1&1&1&1&1&0&0&0&0&0&0&0\\ \noalign{\medskip}1
&1&1&1&1&1&1&0&0&0&0&0&0&0\\ \noalign{\medskip}1&1&1&1&1&1&1&0&0&0&0&0
&0&0\end {array} \right] 
\end{equation}
Note also, that the  
basis functions from the first sector are mirror images of functions from the second sector.
The division of functions into sectors for other $L$ values is similar. The matrix $W_{kl}$ also has 
antidiagonal block form.


\begin{thebibliography}{99}
%
\bibitem{Gu10} S.-J. Gu, 
{\em Fidelity Approach to Quantum Phase Transitions}, 
Int. J. Mod. Phys. B {\bf 24}, 4371 (2010).
%
\bibitem{Wang15}Lei Wang, Ye-Hua Liu, Jakub Imri{\v s}ka, Ping Nang Ma, and Matthias Troyer,
{\em Fidelity Susceptibility Made Simple: A Unified Quantum Monte Carlo Approach},
Phys. Rev. X {\bf 5}, 031007 (2015), DOI: 10.1103/PhysRevX.5.031007.
%
\bibitem{Sun15} G.~Sun, A.K.~Kolezhuk, and T.~Vekua,
{\em Fidelity at Berezinskii-Kosterlitz-Thouless quantum phase transitions},
Phys. Rev. B {\bf 91}, 014418 (2015).
%
\bibitem{Cincio}
 Łukasz Cincio, Marek M. Rams, Jacek Dziarmaga, Wojciech H. Żurek, 
 {\em Universal shift of the fidelity susceptibility peak away from 
 the critical point of the Berezinskii-Kosterlitz-Thouless quantum phase transition},
 Phys. Rev. B {\bf100}, 081108 (2019).
%
\bibitem{Schwandt} David Schwandt, Fabien Alet, and Sylvain Capponi,
{\em Quantum Monte Carlo Simulations of Fidelity at Magnetic Quantum Phase Transitions},
Phys. Rev. Lett. {\bf 103}, 170501 (2009). 
%
\bibitem{Eggert96} Sebastian Eggert, 
{\em Numerical evidence for multiplicative logarithmic corrections from marginal operators}, 
Phys. Rev. B {\bf 54}, R9612 (1996).
%
\bibitem{AffleckWhite92} Steven R. White, Ian Affleck, 
{\em Dimerization and incommensurate spiral spin correlations 
in the zigzag spin chain: Analogies to the Kondo lattice}, 
Phys. Rev. B {\bf 54}, 9862 (1996).
%
\bibitem{Sandv10} Anders W. Sandvik,
{\em Computational Studies of Quantum Spin Systems},
 arXiv:1101.3281v1.
%
\bibitem{Oguchi89} T. Oguchi, H. Kitatani, 
{\em Theory of Resonating Valence Bond Quantum spin system},
J. Phys. Soc. Jpn. {\bf 58}, 1403 (1989).
%
\bibitem{Sandv06} K. S. D. Beach, A. W. Sandvik, 
{\em Some formal results for the valence bond basis},
Nucl. Phys. B {\bf 750}, 142 (2006). 
%
\bibitem{Sandv11} Ying Tang, Anders W. Sandvik, and Christopher L. Henley,
{\em Properties of resonating-valence-bond spin liquids and critical dimer models},
Phys. Rev. B {\bf 84}, 174427 (2011). 
%
\bibitem{Schu} Norbert Schultka and Efstratios Manousakis,
{\em Finite-size scaling in two-dimensional superfluids},
Phys. Rev. B {\bf 49}, 12071 (1994).
%
\bibitem{Kosterlitz} J. M.  Kosterlitz,
{\em The critical properties of the two-dimensional XY model},
Solid State Phys. {\bf 7}, 1046 (1974).
%
\bibitem{Polk15} 	
A. Polkovnikov, V. Gritsev,
{\em Universal Dynamics Near Quantum Critical Points}, hapter 3
in {\em Understanding Quantum Phase Transitions}, 
ed. L. Carr, Taylor \& Francis, Boca Raton, 2010, ISBN 978-1-4398-0251-9.
%
\bibitem{Shi06} Shi-Liang Zhu,
{\em Scaling of Geometric Phases Close to the Quantum Phase Transition in the XY Spin Chain},
Phys. Rev. Lett. {\bf96}, 077206 (2006).
%
\bibitem{Barber} M. N. Barber, 
{\em Finite Size Scaling} in {\em Phase Transitions and Critical Phenomena}, Eds. C. Domb and J. L. Lebowitz,
Academic, 1983, Vol. 8.
\end{thebibliography}
\end{document}